\documentstyle[12pt]{article}
\newcommand{\be}{\begin{equation}}
\newcommand{\ee}{\end{equation}}

\begin{document}
\begin{center}

{\bf NEW TRENDS IN SOLAR-NEUTRINO PHYSICS}\\
Proceedings of 4th International Topical Workshop \\
ed's V.Berezinsky and G.Fiorentini\\
Laboratori Nazionali del Gran Sasso, May 2 - 4 1996\\
\end{center}
\section{Introduction}

	These are the Proceedings of the 4th discussion workshop in the 
Gran Sasso Laboratory. The structure of this workshop was as before: most of 
the time was allocated to the review talks and to the discussions. There were 
discussion sessions and short original presentations for each topic as well as 
the general discussion session at the end of the workshop.

The workshop was focused on three topics:
\begin{itemize}
\item[(i)]	
Solar opacity and collective plasma effects,
\item[(ii)]	
Helioseismological constraints on the SSM's and neutrino fluxes,
\item[(iii)]	
Time variation of solar neutrino flux, its origin and possible
correlation with solar-surface phenomena.
\end{itemize}
	
The topic (i) was presented by two talks:  C. Iglesias described the
"classical" Livermore calculations of solar opacity. R. Bingham presented the
calculations of the group, V. Tsytovich, R. Bingham, U. de Angelis, A. Forlani
and M. Occorsio for collective plasma effects. The total change of opacity due
to collective plasma effects is predicted to be $ -9.3\%$.  C.Iglesias and 
F. Rogers agreed with significance of some effects. It was understood that 
variation of opacity within $10\%$ does not affect drastically predictions for 
both solar neutrino fluxes and helioseismology. In particular, it cannot solve 
the solar-neutrino problem (see contribution by B. Ricci). \\*[1mm]

Helioseismological data, topic (ii), are in excellent agreement with SSM's down
to distance $0.1 R_{\odot}$ . This region covers the site of production of 
pp-neutrinos and it is not far away of production of $^8B$ and $^7Be$ 
neutrinos. It seems rather impossible to predict neutrino fluxes using only 
helioseismological data. It is better to speak about helioseismologically 
constrained SSM's and their predictions for neutrino fluxes. Naturally these 
predictions are those of SSM's.\\*[1mm]

	Much attention was given to topic (iii). The Homestake flux is 
consistent with both constant and time variable fluxes (M. Lissia and 
T.Stanev). The best proof for time variability comes now from correlation 
phenomena, if they are really observed. Most time and attention were given to 
the predictions for correlation supported by the solar models. The great 
help for understanding this problem and for discussions was due to excellent 
talk by P.Hoyng.

	Finally, the neutrino oscillations were discussed.\\*[1mm]

	We want to express our gratitude to the speakers and to many people who
actively participated in the discussions. The meeting would be impossible
without the great personal support of the Director of L.N.G.S., Prof. P.
Monacelli.

	We acknowledge the tremendous contribution of the scientific secretary,
Dr. V. Fantozzi, to the organization and to the work of this meeting.

	Many thanks are to the secretaries Ms. E. Giusti, Ms. M. De Filippo and Ms. P.
Fabbri, for their excellent work.\\*[2mm]
V. Berezinsky,\\*[1mm]								
G. Fiorentini\\

\section{Contents}
Introduction\\
V. Berezinsky, G. Fiorentini\\*[1mm]
Solar neutrino observational data\\ 
T. Kirsten\\*[1mm]
Solar modelling and critical study of observable predictions \\
S. Turck-Chieze\\*[1mm]
Helioseismology, solar models and solar neutrinos\\ 
J. Christensen-Dalsgaard\\*[1mm]
Helioseismology, p-mode inversion and solar-core quasi-rigid rotation \\
E. Fossat\\*[1mm]
Uncertainties in solar opacities\\ 
C. Iglesias\\*[1mm]
Collective plasma processes in the solar interior and the problem of the solar
neutrino deficit \\
R. Bingham\\*[1mm]	
Solar opacity, neutrino signals and helioseismology \\
B. Ricci\\*[1mm]
The evolution of the large-scale magnetic field of the sun \\
P. Hoyng\\*[1mm]
Time variation of solar-neutrino flux \\
M. Lissia\\*[1mm]
Time variations of the Homestake n  flux: an update ('96) \\
T. Stanev\\*[1mm]
Neutrino magnetic moments and time variations of the neutrino flux\\ 
E. Akhmedov\\*[1mm]
On the possibility to measure neutrino magnetic moment down to
$10^{-11}~ \mu_B$\\ 
R. Bernabei\\*[1mm]
Resonance flavor conversion and time variations of the solar neutrino flux
\\ 
A. Smirnov\\*[1mm]
Solar neutrino flux uncertainties and vacuum oscillation and 
MSW ($\nu_e \to \nu_s$) solutions\\ 
S.T. Petcov\\*[1mm]	
Comments on oscillation solutions: why and how? \\
G. Fiorentini	\\*[1mm]
Future solar neutrino experiments \\
F. von Feilitzsch\\*[3mm]
\begin{center}
{\bf Short Presentations during the Discussion Sessions}
\end{center}	
Neutrino and gravity waves\\
E. Schatzman\\*[1mm]
Seismological constraints on the solar neutrino flux \\
I.W. Roxburgh\\*[1mm]
On the new data set from Homestake experiment \\
S. Massetti\\*[1mm]
The effect of the random matter density perturbations on the MSW solution to
the solar neutrino problem \\
H. Nunokawa\\*[1mm]	
Solar neutrino problem and Planck scale physics \\
Z. Berezhiani\\*[1mm]
HELLAZ can detect $\bar{\nu}_e$\\
T. Ypsilantis\\*[1mm]
Concluding remarks \\
E. Fiorini\\*[1mm]

\section{\bf List of libraries where the Proceedings are available}
CERN, Frascati, ICTP, DESY, Stanford, FNAL,
UNICAMP and Sao Paulo University (Brazil), Sudbury Neutrino Observatory and 
McGill University (Canada), University of Copenhagen (Denmark), University 
of Helsinki (Finland), Saclay, Meudon Observatory, Nice University 
and College de France (France), Technical University of Munich,
Max-Planck-Institute for Nuclear Physics (Heidelberg) and Munich 
Max-Planck- Institute for Physik (Germany), TATA Institute (India),
Technion and Hebrew University (Israel), Bologna University, Cagliari 
University, Ferrara University, Florence University, Milano University,
Napoli University, Padova University, Rome 1 and 2 
Universities, Torino University, ICPT and SISSA (Italy), Kyoto Institute 
for Theoretical Physics, Institute for Cosmic Ray Research, KEK (Japan)
INR, ITEP, TH of Valencia University 
(Spain), Berkeley,
MIT, University of Chicago, FNAL, University of Maryland, Princeton, 
University of Pennsylvania (USA).
\end{document}